\documentclass[pdftex,twocolumn,epjc3]{svjour3}

\pdfoutput=1

\RequirePackage[T1]{fontenc}
\smartqed
 
\RequirePackage{graphicx}
\RequirePackage{mathptmx}
\RequirePackage{atlasphysics}
\RequirePackage[numbers,sort&compress]{natbib}
\RequirePackage[colorlinks,citecolor=blue,urlcolor=blue,linkcolor=blue]{hyperref}
\usepackage{epstopdf}
\usepackage{subfigure}
\usepackage{lineno}
\usepackage{amsmath}

\def\ee{\ensuremath{e^{\pm}e^{\pm}}}
\def\emu{\ensuremath{e^{\pm}\mu^{\pm}}}
\def\mm{\ensuremath{\mu^{\pm}\mu^{\pm}}}
\def\ll{\ensuremath{\ell^{\pm}\ell^{\pm}}}

\usepackage{preprintcover}
\PreprintCoverPaperTitle{Search for doubly-charged Higgs bosons in like-sign dilepton final states at $\sqrt{s} = 7$~TeV with the ATLAS detector}
\PreprintIdNumber{CERN-PH-EP-2012-265}
\PreprintCoverAbstract{
A search for doubly-charged Higgs bosons decaying to pairs of electrons and/or muons is presented.  The search is performed using a data sample corresponding to an integrated luminosity of 4.7 \ifb\ of $pp$ collisions at $\sqrt{s}$ = 7 TeV collected by the ATLAS detector at the LHC.  Pairs of prompt, isolated, high-$\pt$ leptons with the same electric charge ($\ee$, $\emu$, $\mm$) are selected, and their invariant mass distribution is searched for a narrow resonance.  No significant excess over Standard Model background expectations is observed, and limits are placed on the cross section times branching ratio for pair production of doubly-charged Higgs bosons.  
The masses of doubly-charged Higgs bosons are constrained depending on the branching ratio into these leptonic final states.
Assuming pair production, coupling to left-handed fermions, and a branching ratio of 100\% for each final state, masses below 409~GeV, 375~GeV, and 398~GeV are excluded for $\ee$, $\emu$, and $\mm$, respectively.
}
\PreprintJournalName{Eur. Phys. J. C}

\newcommand{\hpp}{H^{\pm\pm}}
\newcommand{\hmm}{H^{\mp\mp}}
 
\journalname{Eur. Phys. J. C}

\DeclareGraphicsExtensions{.pdf}

\begin{document}
\sloppy

\title{Search for doubly-charged Higgs bosons in like-sign dilepton final states at $\sqrt{s} = 7$~TeV with the ATLAS detector}

\author{The ATLAS Collaboration}

\institute{CERN, 1211 Geneva 23, Switzerland}

\date{}  

\maketitle

\begin{abstract}
A search for doubly-charged Higgs bosons decaying to pairs of electrons and/or muons is presented.  The search is performed using a data sample corresponding to an integrated luminosity of 4.7 \ifb\ of $pp$ collisions at $\sqrt{s}$ = 7 TeV collected by the ATLAS detector at the LHC.  Pairs of prompt, isolated, high-$\pt$ leptons with the same electric charge ($\ee$, $\emu$, $\mm$) are selected, and their invariant mass distribution is searched for a narrow resonance.  No significant excess over Standard Model background expectations is observed, and limits are placed on the cross section times branching ratio for pair production of doubly-charged Higgs bosons.  
The masses of doubly-charged Higgs bosons are constrained depending on the branching ratio into these leptonic final states.
Assuming pair production, coupling to left-handed fermions, and a branching ratio of 100\% for each final state, masses below 409~GeV, 375~GeV, and 398~GeV are excluded for $\ee$, $\emu$, and $\mm$, respectively.
\end{abstract}


Several extensions of the Standard Model (SM) predict the existence of
doubly-charged Higgs bosons ($\hpp$) as part of an extended Higgs sector
where a triplet of Higgs bosons is present~\cite{triplet1,triplet2,hpp_triplet,hpp_little}. 
In addition to the neutral Higgs boson, charged and doubly-charged Higgs bosons
are part of this triplet. The origin of neutrino masses and mixing can be attributed to this triplet,
which can couple to Higgs and lepton doublets. Dirac neutrino mass terms are generated via the Yukawa couplings of the 
left-handed leptons to the Higgs triplet. This is commonly known as the ``type II seesaw'' mechanism~\cite{seesaw1,seesaw2,seesaw3,seesaw4}.
A $\hpp$-like particle can also occur as a singlet as proposed in the Zee-Babu model~\cite{ZBM1, ZBM2, ZBM3}, 
where it is postulated for the purpose of generating Majorana neutrino masses.

The main production mechanisms of $\hpp$ bosons at hadron colliders are pair production via an $s$-channel $Z$-boson or photon exchange
and associated production with a $H^\pm$ boson via a $W$-boson exchange. The latter process depends on the mass of the $H^\pm$ 
boson, which is generally unknown, and is not considered here. Doubly-charged Higgs bosons can couple to either left-handed or right-handed fermions. In left-right symmetric models~\cite{LRSM1,LRSM2,LRSM3,LRSM4}, the two cases are distinguished and denoted $\hpp_{L}$ and $\hpp_{R}$. The cross section for $H^{++}_{L}H^{--}_L$ production is about 2.5 times larger than that for $H^{++}_{R}H^{--}_R$ production due to different couplings to the $Z$ boson~\cite{spira}. 

The $\hpp$ boson may decay to a pair of like-sign leptons whose invariant mass is consistent with the mass of the $\hpp$ boson, denoted $m(\hpp)$. The partial decay width to leptons is given by 
\begin{equation}
\Gamma(\hpp \to \ell^\pm\ell'^\pm) = k\frac{h_{\ell\ell'}^2}{16 \pi}m(\hpp),
\end{equation}
where $k=2$ if both leptons have the same flavour ($\ell=\ell'$) and $k=1$ if they have a different flavour. 
The factor $h_{\ell\ell'}$ is the coupling parameter. 
Only prompt decays of the $\hpp$ boson ($c\tau < 10$~$\mu$m) are considered in this letter, corresponding to $h_{\ell\ell'} > 3\times 10^{-6}$ for $m(\hpp) > 50$~GeV.

Direct searches for $\hpp$ have been performed by experiments at the LEP, HERA, Tevatron, and LHC colliders~\cite{pdg_2012, 2011SSmm, cmsHpp}. The mass limits obtained for the $\hpp$ bosons depend on the branching ratio (BR) assumed.  
The most stringent limits were set by the CMS Collaboration~\cite{cmsHpp}, excluding at 95\% confidence level (CL) masses below 382~GeV, 391~GeV, and 395~GeV for $\ee$, $\mm$, and $\emu$ final states, respectively, assuming left-handed couplings and a 100\% branching ratio to each final state.
Low-energy constraints exist on the product of some of the coupling parameters. The 90\% CL upper limit on the muon decay branching ratio BR$(\mu^+ \to e^+ e^- e^+)<1.0\times 10^{-12}$ provides the tightest constraint, $h_{ee}h_{e\mu}/m(\hpp)^2<4.7 \times 10^{-11}$~GeV$^{-2}$~\cite{mudecay},
but the individual couplings are not constrained. More details 
on constraints from non-collider experiments can be found in Ref.~\cite{theory_hpphmm}.

This letter presents a search for a $\hpp$ boson decaying to pairs of electrons and/or muons with the same electric charge ($\hpp\to\ee$, $\hpp\to\mm$, and $\hpp\to\emu$).  
The search is performed using a data sample corresponding to $4.7\pm 0.2$ $\ifb$~\cite{lumipaper, atlaslumi2011} of integrated luminosity (${\cal L}$) collected by the ATLAS detector~\cite{atlas}.  
The event selection and background estimates follow those used by the inclusive search for
anomalous production of like-sign leptons described in
Ref.~\cite{inclpaper} and are summarized only briefly here.

Events are collected using single-lepton triggers with transverse momentum ($\pt$)
thresholds of 18~GeV for muons and either 20~GeV or 22~GeV for electrons, depending on the running period. 
To ensure no efficiency loss for electrons with very high $\pt$, a trigger with a threshold of 45~GeV and looser requirements for the electron identification is also used.
Electrons are identified as showers in the electromagnetic calorimeter matched to a track in the inner detector using the
{\it tight} criteria described in Ref.~\cite{epaper}. Muons
are reconstructed from combined tracks in the inner tracking system
and the muon detector~\cite{atlas}. 
Leptons must have a transverse momentum above 20~GeV,
be in the angular range covered by the inner tracking system, 
be well isolated, and have impact parameters consistent with
originating from the primary event vertex. In pairs where
the higher-$\pt$ lepton is an electron, it is required to have
$\pt>25$~GeV.  All pairs of electrons or muons with the same electric charge are considered,
so more than one lepton pair may be reconstructed per event.
The invariant mass of the lepton pair must be larger than 15~GeV, and
for $\ee$ the region close to the $Z$-boson mass
(70 GeV $< m(\ee) < 110$ GeV) is excluded due to a large background from $Z \to e^+e^-$ events with an electron charge misidentification, as described below.

SM backgrounds in this search are divided into three categories:
prompt, non-prompt, and conversion sources.  The
dominant source of prompt like-sign lepton pairs is $WZ$ production,
and smaller sources include $ZZ$, like-sign $WW$,
$t\bar{t}W$, and $t\bar{t}Z$ production. The expected prompt background contribution 
is derived from MC simulation normalised to cross-section calculations
performed at next-to-leading order in the strong coupling constant. The background due to
non-prompt leptons comes primarily from semileptonic $b$- and $c$-hadron decays for muons,
while for electrons, an additional background comes from charged pions that shower early in the calorimeter
and neutral pions decaying to two photons, where one of the photons converts to an $e^+e^-$ pair.
This background is derived by extrapolating from data regions where
one or both leptons fail part of the selection criteria. The
extrapolation factors are derived in events dominated by non-prompt
leptons and are dependent on the $\pt$ and pseudorapidity ($\eta$) of the leptons.
The origin of the background due to photon conversions, important for the electron channels, includes $W\gamma$ and $Z\gamma$ production and cases where one of the electrons in an $e^+e^-$ pair is reconstructed with the wrong charge (electron charge flip) after radiating a photon that subsequently converts.
The background from conversions is derived using MC simulation, where the electron
charge-flip rate has been corrected based on comparison between data
and simulation for $Z\to e e$ events. The charge-flip background due to the finite curvature resolution 
is small compared to that from conversion sources.

Several control regions are selected
to validate the background estimates.  In all cases, the dilepton mass
spectrum observed in data agrees with the background prediction within
the systematic uncertainties~\cite{inclpaper}.

The $\hpp$ process is simulated using {\sc pythia8}~\cite{pythia} to estimate
the acceptance and efficiency for reconstructing $\hpp$ bosons with masses between 50~GeV and
1000~GeV. The kinematic properties of $\hpp_L$ and $\hpp_R$ bosons are identical; only their cross sections differ.
Windows in the reconstructed dilepton mass are defined to ensure optimal signal sensitivity across the full mass range.   
In each window, the number of observed events is compared to the expected background and signal yields to derive a limit on the signal contribution.
The expected observable width of the $\hpp$ resonance is dominated by detector
resolution effects.  The detector resolution of the momentum measurement is roughly 1.2--1.8\% for high-\pt{} electrons, depending
on \eta, and for muons ranges from about 2\% for $\pt=20$~GeV to 10\% for $\pt=1$~TeV.
For the $\ee$ final state, lepton pairs with a mass within $\pm 4\%$ of the tested mass value are selected.  For $\mm$ and $\emu$, 
the window size is $\pm (6 + 0.007\times m(\hpp)/{\rm GeV})\%$ around the tested mass to account for the coarser resolution. With the selection described above, the
acceptance times efficiency to reconstruct a $\hpp$ boson in these mass windows is about 27\% for $\ee$, 36\% for $\emu$, and 43\% for $\mm$ for $m(\hpp)=100$~GeV. For $m(\hpp)=400$~GeV it is about 50\% for all three final states. 

Among the dominant systematic uncertainties on the background yields are the $\pm12\%$ uncertainties on the $WZ$ and $ZZ$ cross sections.  
The cross-section uncertainties on the $t\bar{t}W$, $t\bar{t}Z$, and like-sign $WW$ processes are $\pm50\%$, but the contribution from these backgrounds is small.
The non-prompt and conversion background uncertainties are also significant: at low mass, they are about $\pm40\%$ and stem mainly from the uncertainty in the extrapolation factor used for the non-prompt estimate and uncertainty in the rate of photon-to-electron conversions.  At higher mass, the uncertainties on all backgrounds are dominated by the limited size of MC or data samples used in their estimates.
Smaller uncertainties arise from the lepton identification, isolation, and trigger efficiencies, which also apply to the $\hpp$ signal acceptance. For the signal, an additional uncertainty of $\pm1.6\%$ is estimated from the parton distribution functions by using the uncertainties provided by the MSTW 2008 90\% CL set~\cite{mstw2008} added in quadrature to the difference between the central value of this set and the CTEQ6L PDF set.

The dilepton mass distribution observed in data is shown for the $\ee$, $\mm$, and $\emu$ channels in Fig.~\ref{fig:hppmass} and is compared to the background expectation and four hypothetical $\hpp$ signals normalised to their respective cross sections (assuming a branching ratio to the given lepton flavour of $100\%$). The data show no clear peak structure and agree well with the background estimate in all three channels. 

\begin{figure}[htbp]
\begin{center}
\subfigure[] {
  \includegraphics[width=1.0\columnwidth]{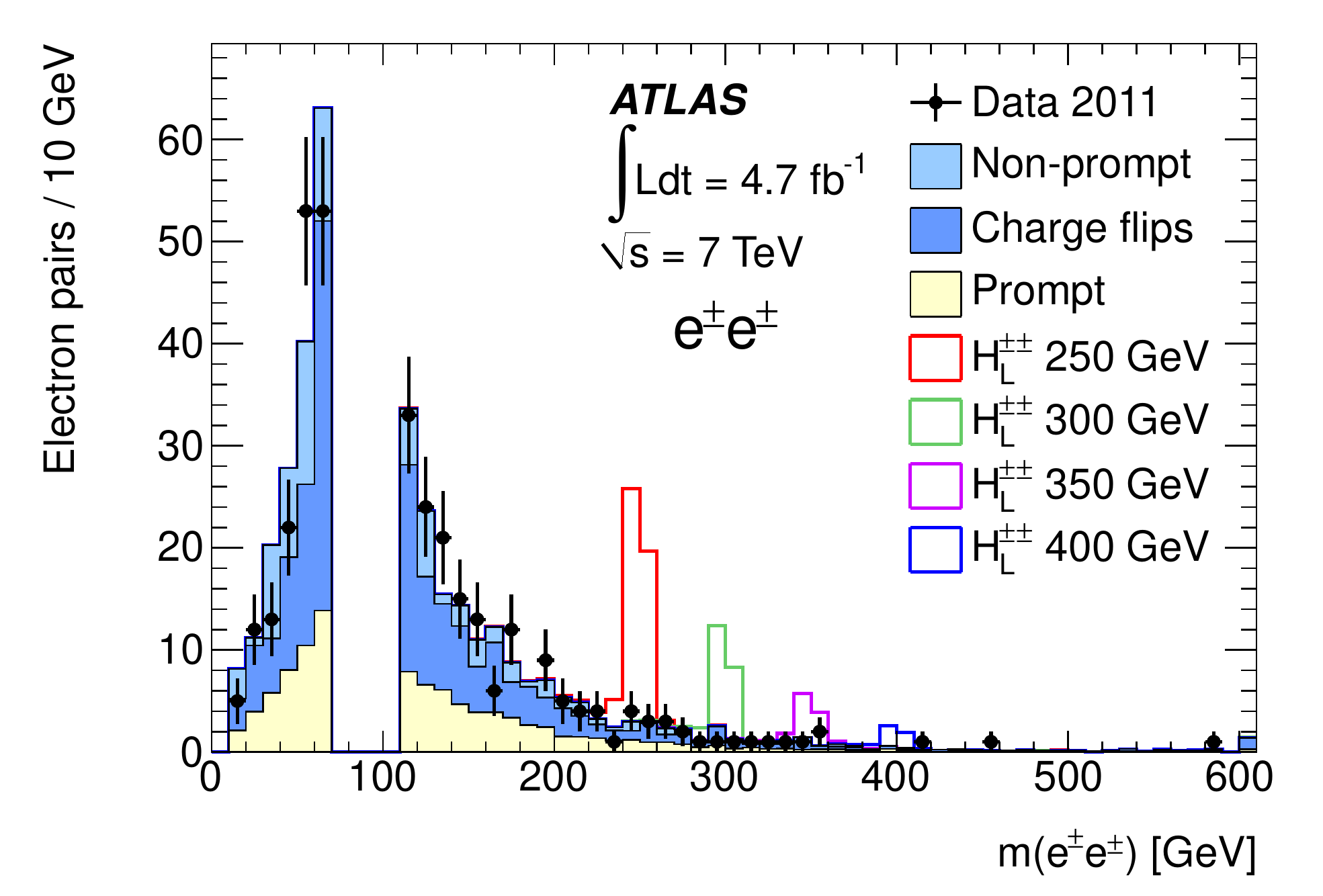}
}
\subfigure[] {
 \includegraphics[width=1.0\columnwidth]{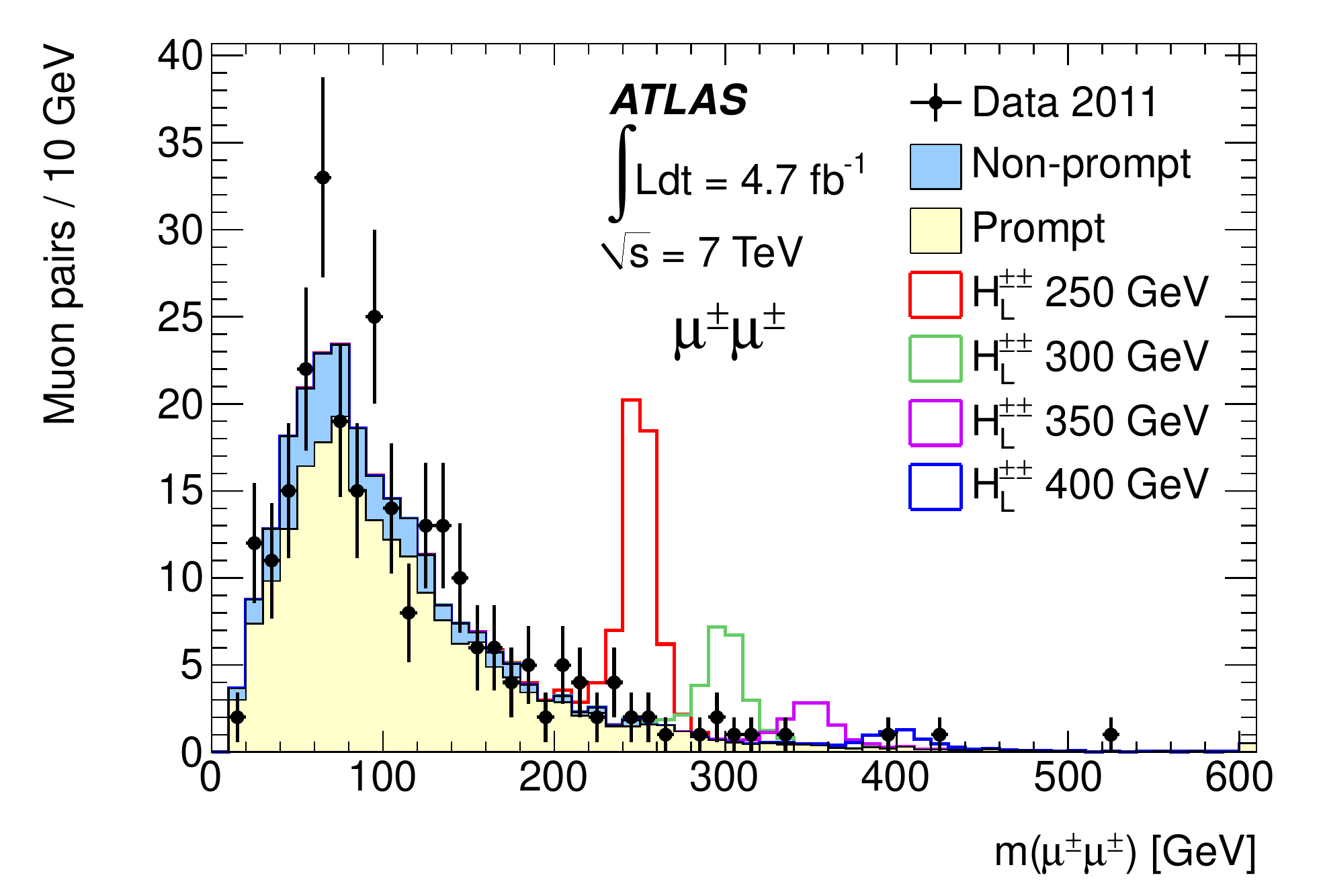}
}
\subfigure[] {
  \includegraphics[width=1.0\columnwidth]{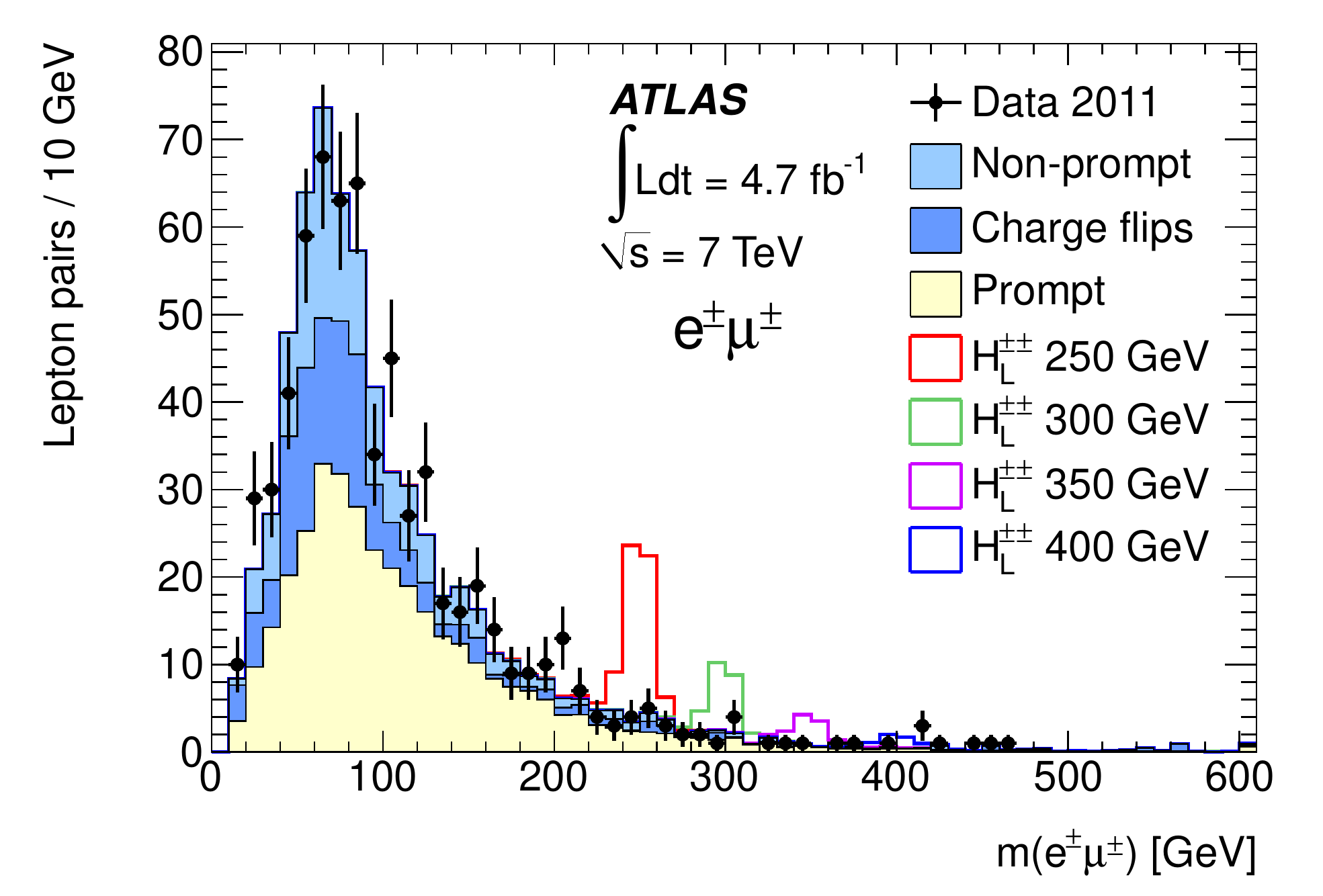}
}
\caption{\label{fig:hppmass} Invariant mass distributions for (a) $\ee$, (b) $\mm$, and (c) $\emu$
pairs passing the full event selection. The data are shown as filled circles.  The stacked histograms represent the backgrounds composed of pairs of prompt leptons from SM processes, pairs with at least one non-prompt lepton, and for the electron channels, backgrounds arising from charge misidentification and conversions.  The open histograms show the expected signal from simulated $\hpp_L$ samples, assuming a 100\% branching ratio to the decay channel considered and coupling to left-handed fermions. Lepton pairs in the \ee\ channel with an invariant mass between 70~GeV and 110~GeV are excluded because of the larger background from charge misidentification in $Z\to e^{\pm}e^{\mp}$ decays. The last bin is an overflow bin.} 
\end{center}
\end{figure}

A limit on the number of lepton pairs originating from $\hpp$ bosons ($N^{\rm rec}$) in each mass
window is derived using a CLs technique~\cite{CLs}. It is converted to a limit on the cross section times branching ratio for doubly-charged Higgs production using the acceptance times efficiency values derived from MC simulation. Since this analysis counts lepton pairs and each event contains two $\hpp$ bosons, the cross section times branching ratio for pair production is given by 
\begin{multline}
\sigma(pp\to \hpp\hmm) \times BR(\hpp \to \ell^\pm \ell'^\pm) = \\
\frac{N^{\rm rec}(\ell^\pm\ell'^\pm)}{2\times A \times \varepsilon \times {\cal L}}, 
\end{multline}
where $A\times\varepsilon$ is the acceptance times efficiency to detect a lepton pair from $\hpp$ decay within a given mass window. The integrated luminosity $\cal L$ is 4.7 $\ifb$.  

The 95\% CL expected and observed upper limits on the cross section times branching ratio as a function of the $\hpp$ boson mass are shown in Fig.~\ref{fig:limit}. The expected limit is determined as the median outcome of simulated pseudo-experiments in the absence of any signal. Also shown are the theoretical cross sections calculated at next-to-leading order (NLO) for $\hpp$ production with left- and right-handed couplings~\cite{spira}. The uncertainty on these cross sections is $\pm 10\%$ due to scale dependence in the NLO calculation, parton distribution function uncertainties, and neglecting higher-order electroweak corrections.

\begin{figure}[htbp]
\begin{center}
\subfigure[]{
  \includegraphics[width=1.0\columnwidth]{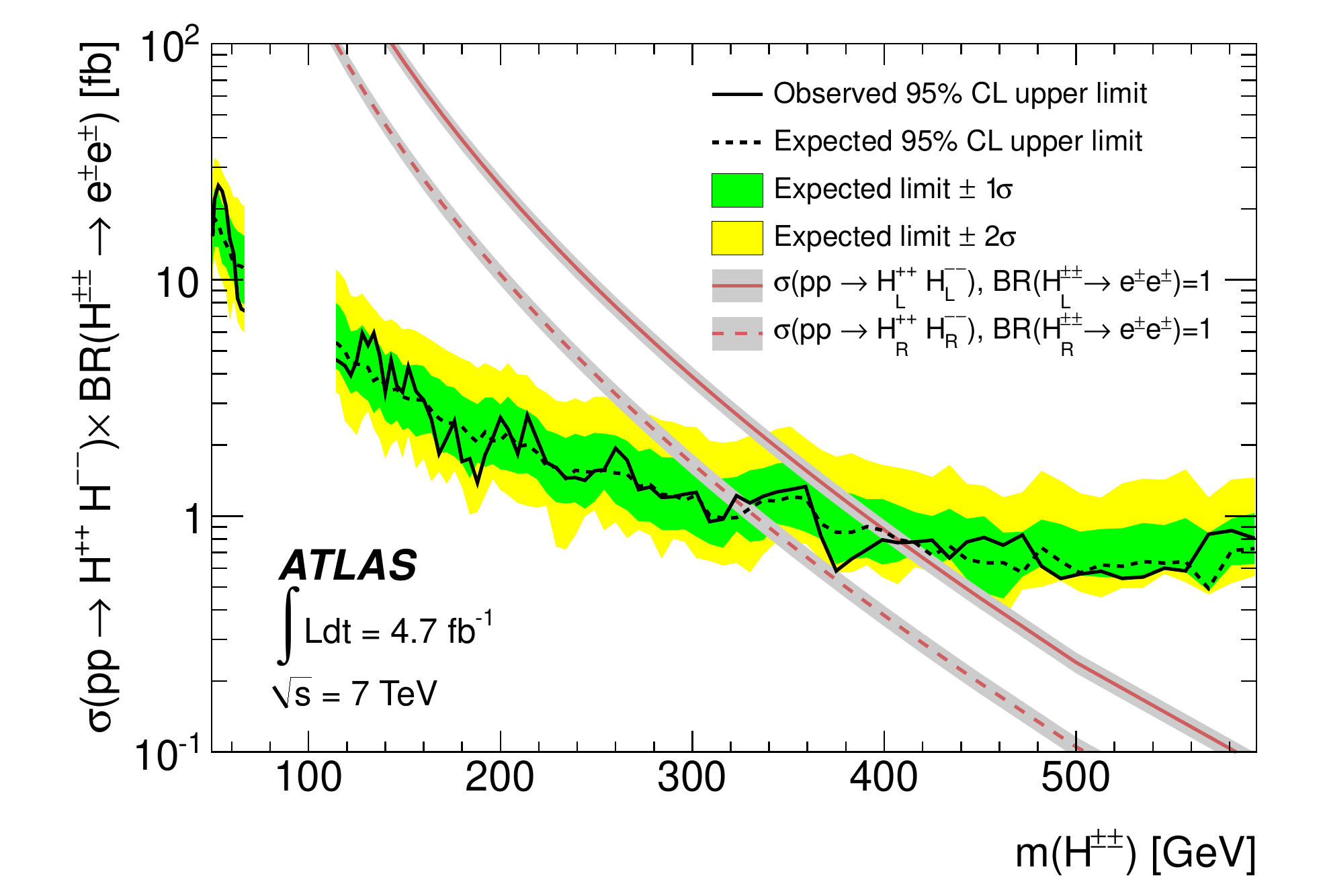}
}
\subfigure[]{
  \includegraphics[width=1.0\columnwidth]{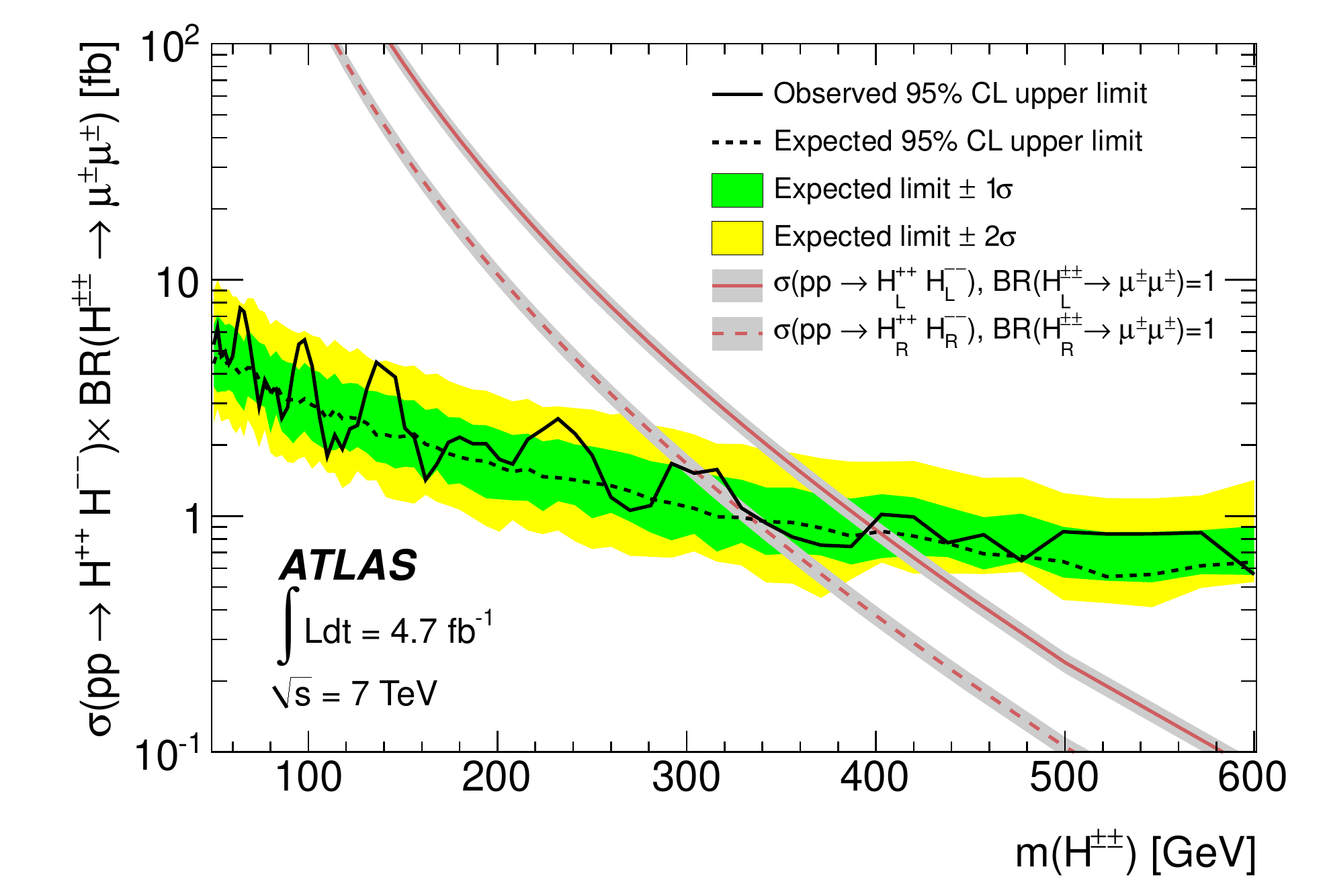}
}
\subfigure[]{
  \includegraphics[width=1.0\columnwidth]{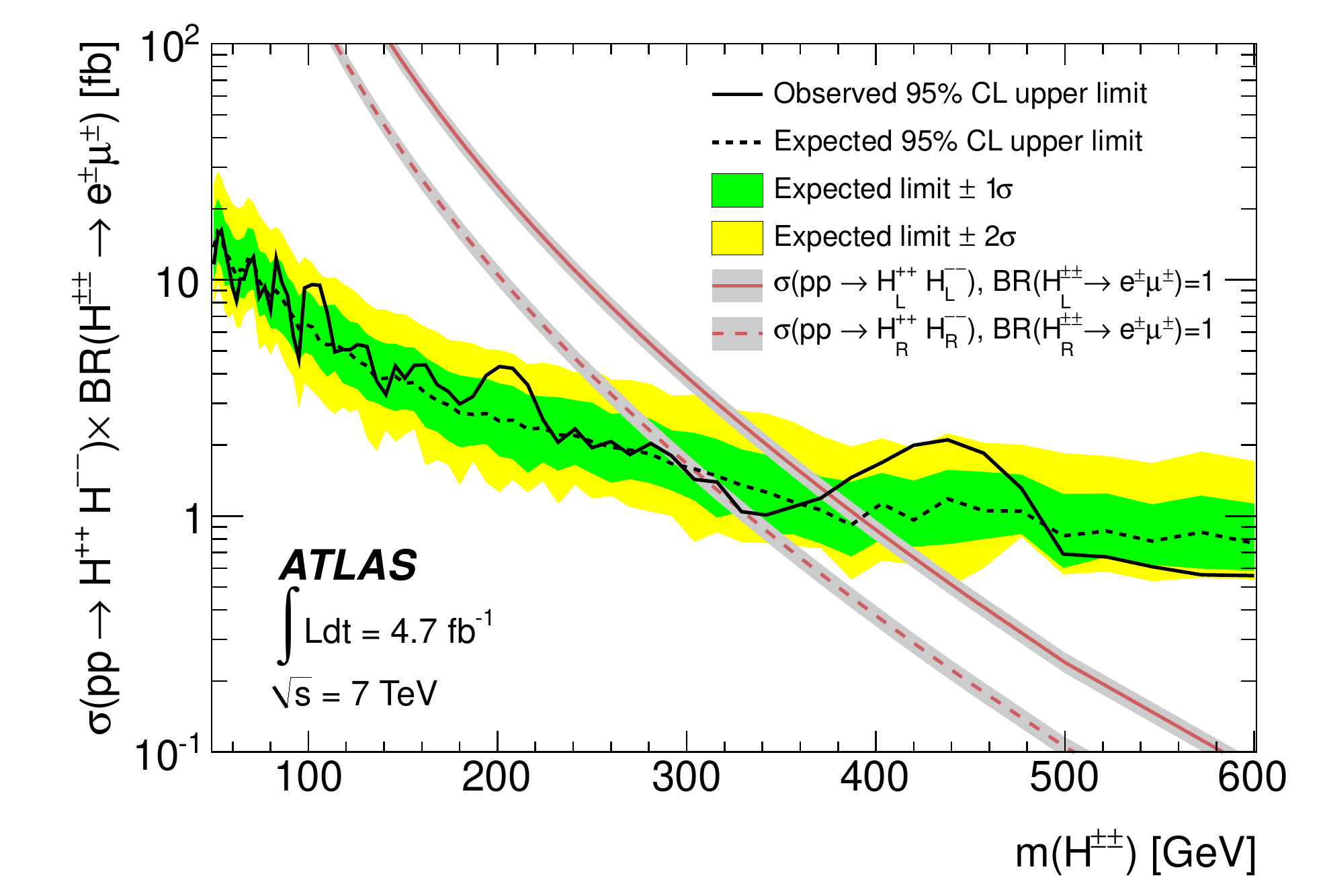}
}
\caption{\label{fig:limit} Upper limit at 95\% CL on the cross section times branching ratio for pair production of $\hpp$ bosons decaying to (a) $\ee$, (b) $\mm$, and (c) $\emu$ pairs.  The observed and median expected limits are shown along with the $1\sigma$ and $2\sigma$ variations in the expected limits.
In the range $70 < m(\hpp) < 110$~GeV, no limit is set in the $\ee$ channel. Also shown are the theoretical
predictions at next-to-leading order for the $pp\to \hpp\hmm$ cross section for $\hpp_L$ and $\hpp_R$ bosons.
The variation from bin to bin in the expected limits is due to fluctuations in the background yields derived from small MC samples.
}
\end{center}
\end{figure}

At low mass, the expected cross-section limits are most stringent for the $\mm$ channel due to the low background levels in this channel. At high mass, the expected $\ee$ and $\mm$ limits are comparable while the $\emu$ limit is about 30\% worse due to the larger background from $WZ$ production. In general the observed and expected limits agree well with each other. The largest deviations of the observed limit from the expected limit are within the 2$\sigma$ uncertainty on the expected limit. The cross-section limits range from 25~fb (in the $\ee$ channel at low mass) to 0.6~fb (in all channels at high mass).

Comparison of the cross-section limits with the theoretical production cross section places constraints on $m(\hpp)$. The lower limits on the $\hpp$ mass at 95\% CL are listed in Table~\ref{tab:mass} for the three final states when BR$(\hpp \to \ell^{\pm}\ell'^{\pm})=100\%$, as well as branching ratios of 33\%, 22\%, and 11\%.  For a democratic scenario where the BR to each pair of lepton flavours is the same, the branching ratio is 22\% for the $\ee$ and $\mm$ final states and 11\% for the $\emu$ final state.
In addition, the same mass limits can be placed on the singlet $H^{\pm\pm}$ in the Zee-Babu model as its production cross sections and decay kinematics are the same as for $H_L^{\pm\pm}$.  Figure~\ref{fig:limitvsbr} shows the mass limits as a function of the branching ratio into each of the three final states.

\begin{table}
\caption{Lower mass limits at 95\% CL on $\hpp$ bosons decaying to $\ee$, $\mm$, or $\emu$ pairs.  Mass limits are derived assuming branching ratios to a given decay mode of 100\%, 33\%, 22\%, or 11\%.  Both expected and observed limits are given.}
\begin{center}
\begin{tabular}{c||c|c||c|c||c|c}
BR$(\hpp_L \to \ell^{\pm}\ell'^{\pm})$ & \multicolumn{6}{c}{95\% CL lower limit on $m(\hpp_L)$ [GeV]} \\[+0.05in]
\hline
 & \multicolumn{2}{c||}{$e^{\pm}e^{\pm}$} & \multicolumn{2}{c||}{$\mu^{\pm}\mu^{\pm}$} & \multicolumn{2}{c}{$e^{\pm}\mu^{\pm}$} \\
 & exp. & obs. & exp. & obs. & exp. & obs.  \\
\hline
100\% & 407 & 409 & 401 & 398 & 392 & 375 \\
33\%  & 318 & 317 & 317 & 290 & 279 & 276 \\
22\%  & 274 & 258 & 282 & 282 & 250 & 253 \\
11\%  & 228 & 212 & 234 & 216 & 206 & 190 \\
\hline
\hline
BR$(\hpp_R \to \ell^{\pm}\ell'^{\pm})$ & \multicolumn{6}{c}{95\% CL lower limit on $m(\hpp_R)$ [GeV]} \\[+0.05in]
\hline
 & \multicolumn{2}{c||}{$e^{\pm}e^{\pm}$} & \multicolumn{2}{c||}{$\mu^{\pm}\mu^{\pm}$} & \multicolumn{2}{c}{$e^{\pm}\mu^{\pm}$} \\
 & exp. & obs. & exp. & obs. & exp. & obs.  \\
\hline
100\% & 329 & 322 & 335 & 306 & 303 & 310 \\
33\%  & 241 & 214 & 247 & 222 & 220 & 195 \\
22\%  & 203 & 199 & 223 & 212 & 194 & 187 \\
11\%  & 160 & 151 & 184 & 176 & 153 & 151 \\
\hline
\end{tabular}
\end{center}
\label{tab:mass}
\end{table}

\begin{figure}[htbp]
\begin{center}
\subfigure[] {
  \includegraphics[width=1.0\columnwidth]{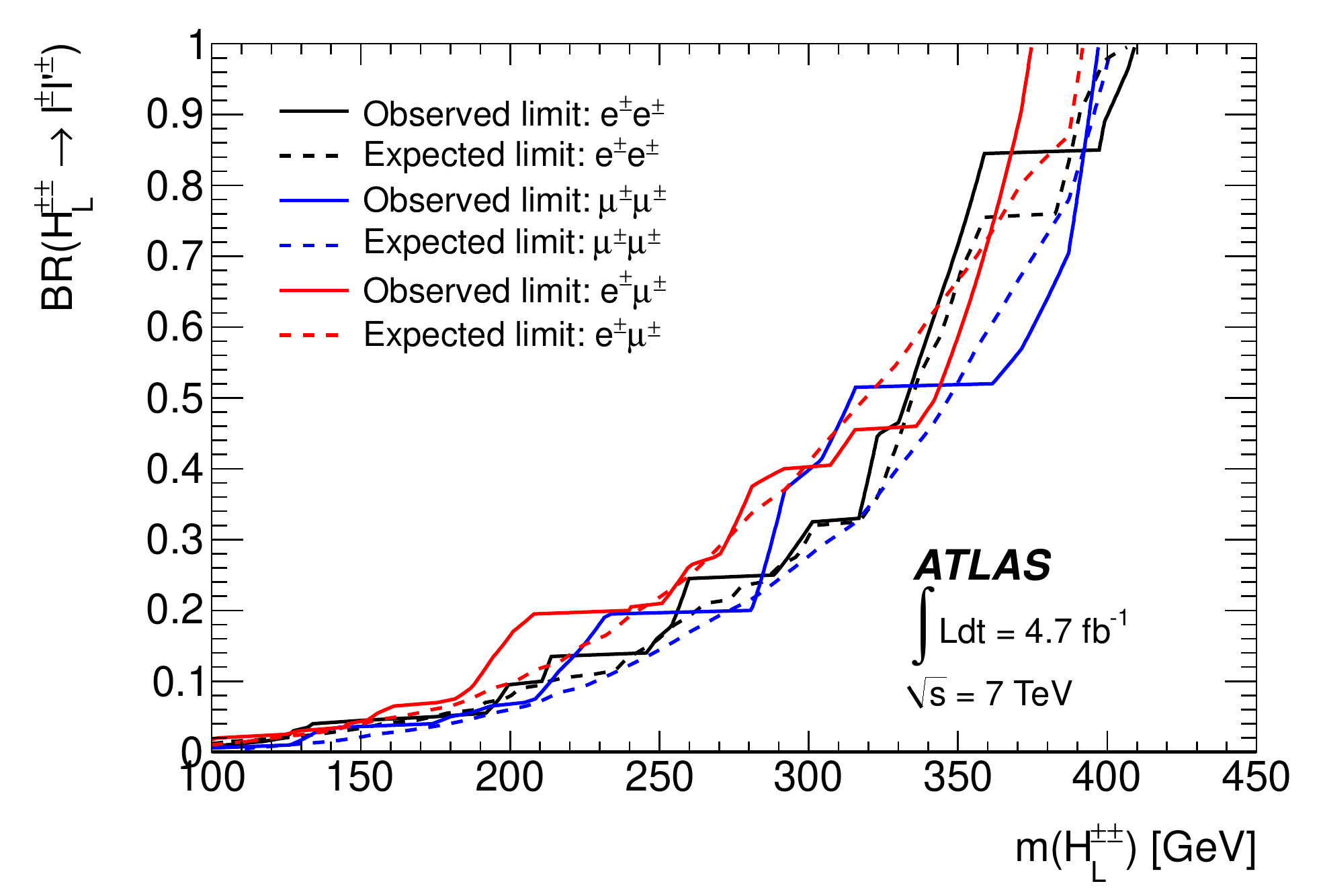}
}
\subfigure[] {
  \includegraphics[width=1.0\columnwidth]{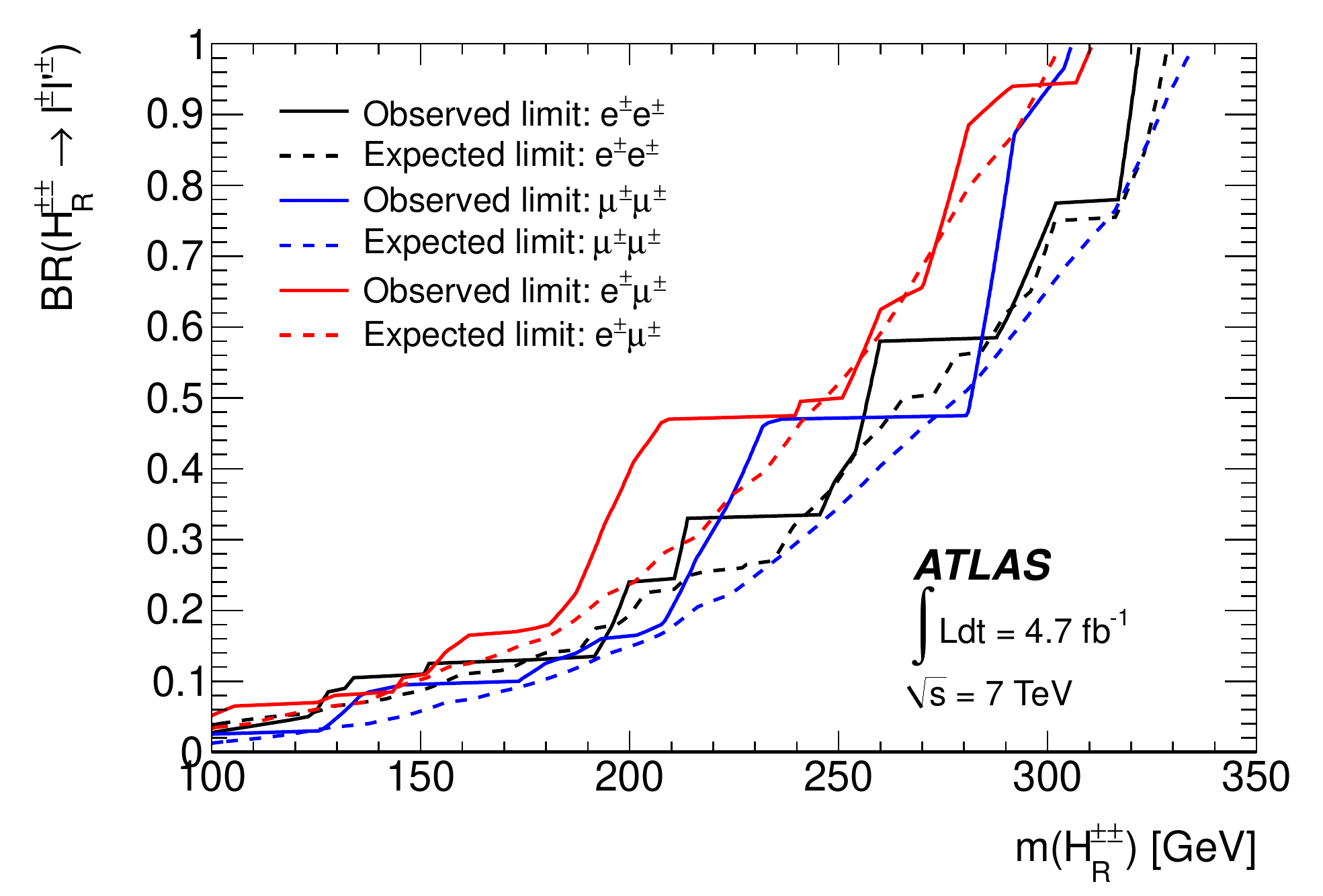}
}
\caption{\label{fig:limitvsbr} The mass limits as a function of the branching ratio for the $\hpp$ decaying to $\ee$, $\emu$, and $\mm$ for (a) $\hpp_L$ and (b) $\hpp_R$ bosons. Shown are both the observed limits (solid lines) and the expected limits (dashed lines).  The stepping behaviour, where the same mass limit is valid for a range of branching ratios, results from fluctuations in the observed cross-section limits shown in Fig.~\ref{fig:limit}.}
\end{center}
\end{figure}

In conclusion, a search for doubly-charged Higgs bosons decaying to $\ee$, $\emu$, or $\mm$ has been performed by searching for a narrow resonance peak in the dilepton mass distribution. 
No such peak was observed in a data sample corresponding to an integrated luminosity of 4.7~fb$^{-1}$ of $pp$ collisions at $\sqrt{s}=7$~TeV recorded by the ATLAS detector at the LHC in 2011. Cross-section limits between 17~fb and 0.6~fb are set depending on the mass of the $\hpp$ boson and the final state. 
Assuming pair production, couplings to left-handed fermions, and a branching ratio of 100\% for each final state, masses below 409~GeV, 398~GeV, and 375~GeV are excluded at 95\% CL for $\ee$, $\mm$, and $\emu$ final states, respectively. Lower mass limits are also set for scenarios with right-handed couplings or smaller branching ratios.  The limits on $\hpp_L$ bosons also apply to the singlet in the Zee-Babu model.

\begin{acknowledgements}

We thank CERN for the very successful operation of the LHC, as well as the
support staff from our institutions without whom ATLAS could not be
operated efficiently.

We acknowledge the support of ANPCyT, Argentina; YerPhI, Armenia; ARC,
Australia; BMWF and FWF, Austria; ANAS, Azerbaijan; SSTC, Belarus; CNPq and FAPESP,
Brazil; NSERC, NRC and CFI, Canada; CERN; CONICYT, Chile; CAS, MOST and NSFC,
China; COLCIENCIAS, Colombia; MSMT CR, MPO CR and VSC CR, Czech Republic;
DNRF, DNSRC and Lundbeck Foundation, Denmark; EPLANET and ERC, European Union;
IN2P3-CNRS, CEA-DSM/IRFU, France; GNSF, Georgia; BMBF, DFG, HGF, MPG and AvH
Foundation, Germany; GSRT, Greece; ISF, MINERVA, GIF, DIP and Benoziyo Center,
Israel; INFN, Italy; MEXT and JSPS, Japan; CNRST, Morocco; FOM and NWO,
Netherlands; BRF and RCN, Norway; MNiSW, Poland; GRICES and FCT, Portugal; MERYS
(MECTS), Romania; MES of Russia and ROSATOM, Russian Federation; JINR; MSTD,
Serbia; MSSR, Slovakia; ARRS and MVZT, Slovenia; DST/NRF, South Africa;
MICINN, Spain; SRC and Wallenberg Foundation, Sweden; SER, SNSF and Cantons of
Bern and Geneva, Switzerland; NSC, Taiwan; TAEK, Turkey; STFC, the Royal
Society and Leverhulme Trust, United Kingdom; DOE and NSF, United States of
America.

The crucial computing support from all WLCG partners is acknowledged
gratefully, in particular from CERN and the ATLAS Tier-1 facilities at
TRIUMF (Canada), NDGF (Denmark, Norway, Sweden), CC-IN2P3 (France),
KIT/GridKA (Germany), INFN-CNAF (Italy), NL-T1 (Netherlands), PIC (Spain),
ASGC (Taiwan), RAL (UK) and BNL (USA) and in the Tier-2 facilities
worldwide.

\end{acknowledgements}

\bibliographystyle{spphys}

\onecolumn
\clearpage
~
\begin{flushleft}
{\Large The ATLAS Collaboration}

\bigskip

G.~Aad$^{\rm 48}$,
T.~Abajyan$^{\rm 21}$,
B.~Abbott$^{\rm 111}$,
J.~Abdallah$^{\rm 12}$,
S.~Abdel~Khalek$^{\rm 115}$,
A.A.~Abdelalim$^{\rm 49}$,
O.~Abdinov$^{\rm 11}$,
R.~Aben$^{\rm 105}$,
B.~Abi$^{\rm 112}$,
M.~Abolins$^{\rm 88}$
% ...
\bigskip

$^{1}$ School of Chemistry and Physics, University of Adelaide, Adelaide, Australia\\
% ...
$^{11}$ Institute of Physics, Azerbaijan Academy of Sciences, Baku, Azerbaijan\\
$^{12}$ Institut de F\'{i}sica d'Altes Energies and Departament de F\'{i}sica de la Universitat Aut\`{o}noma de Barcelona and ICREA, Barcelona, Spain\\
% ...
$^{21}$ Physikalisches Institut, University of Bonn, Bonn, Germany\\
% ...
$^{48}$ Fakult\"{a}t f\"{u}r Mathematik und Physik, Albert-Ludwigs-Universit\"{a}t, Freiburg, Germany\\
$^{49}$ Section de Physique, Universit\'{e} de Gen\`{e}ve, Geneva, Switzerland\\
% ...
$^{88}$ Department of Physics and Astronomy, Michigan State University, East Lansing MI, United States of America\\
% ...
$^{105}$ Nikhef National Institute for Subatomic Physics and University of Amsterdam, Amsterdam, Netherlands\\
% ...
$^{111}$ Homer L. Dodge Department of Physics and Astronomy, University of Oklahoma, Norman OK, United States of America\\
$^{112}$ Department of Physics, Oklahoma State University, Stillwater OK, United States of America\\
% ...
$^{115}$ LAL, Universit\'{e} Paris-Sud and CNRS/IN2P3, Orsay, France\\
% ...
$^{s}$ Also at CPPM, Aix-Marseille Universit\'{e} and CNRS/IN2P3, Marseille, France\\
% ...
\end{flushleft}

\end{document}